\documentclass[aps, amsfonts, nofootinbib,notitlepage, preprintnumbers]{revtex4-1}
\usepackage{hyperref}
\hypersetup{
colorlinks=true,
linkcolor=red,
citecolor=blue,
urlcolor=blue}
\usepackage{epsf,epsfig}
\usepackage{amscd}
\usepackage{amsmath}
\usepackage{subfig}
\usepackage{graphicx}
\usepackage{xcolor}
\usepackage[countmax]{subfloat}
\input xy
\xyoption{all}

\newcommand{\bea}{\begin{eqnarray}}
\newcommand{\eea}{\end{eqnarray}}

\newcommand{\eps}{\epsilon}

\newcommand{\nn}{\nonumber}

\begin{document}

\title{Large $N_c$ gauge theory with  quarks in high representations} 
\author{Thomas D.~Cohen\footnote{{\tt cohen@umd.edu}}} 
\author{Srimoyee Sen\footnote{{\tt srimoyee@umd.edu}}}
\affiliation{Maryland Center for Fundamental Physics,\\ 
Department of Physics,\\ 
University of Maryland, College Park, MD USA}


\begin{abstract}
	
This paper explores a novel tractable regime for ultraviolet-complete quantum field theories---the large $N_c$ limit of non-abelian gauge theories with quarks in high dimensional representations (scaling with $N_c$ faster than $N_c^2$), such as quarks with `a' fundamental indices with $a \ge 3$. A smooth and nontrivial $N_c$ limit can be obtained if $g^2 N_c^{a-1}$ is held fixed instead of the standard 't Hooft coupling $g^2N_c$ as $N_c \rightarrow \infty$ where $g$ is the gauge coupling. SU($N_c$) gauge theories in 3+1 dimensions are not asymptotically free at large $N_c$ when they contain quarks in representations whose dimensions scale faster than $N_c^2$ and hence are not ultraviolet complete.  However, in lower  space-time dimensions  (2+1, 1+1), for any $N_c$,  renormalization group flow for such theories always has a stable ultraviolet fixed point at $g=0$; the theory is thus ultraviolet complete. For the case of massless quarks, the theory has an infrared fixed point. For massive quarks, the theory is confining. The confining scale is parametrically of the order $N_c^{\frac{2-a}{4-d}}$ and is driven to zero at large $N_c$ for theories with $a>2$ and $d<4$ where $d$ is the space time dimension.      	
\end{abstract}
\maketitle

\section{Introduction}

It is challenging to understand strongly coupled nonabelian gauge theories, since a perturbative expansion in the coupling constant is not suitable. This has led researchers to investigate different limits of gauge theories to gain insights. One such limit, the large $N_c$ limit, was proposed by 't Hooft in 1973 \cite{'tHooft:1973jz}. In this limit, the number of colors $N_c$, was taken to infinity, $g \rightarrow 0$, while keeping $g^2N_c$ fixed, where $g$ was the coupling constant. The theory remains strongly coupled since the relevant coupling is not $g^2$ but $g^2N_c$.

In 't Hooft's original analysis, the quarks were in the fundamental representation of SU($N_c$) and the number of flavors was kept constant. This limit has interesting consequences, one of which is the suppression of quark loops, thus the gluodynamics, at leading order, is decoupled from the quark dynamics. On the other hand, it was recognized quite early that the large $N_c$ limit of SU($N_c$) gauge symmetry is not unique: even if the gauge is fixed, one can include fermions in a variety of ways yielding physically distinct large $N_c$ limits. For example, G Veneziano suggested another interesting limit \cite{Veneziano:1979ec} where $N_f\rightarrow\infty$ and $N_c\rightarrow\infty$ keeping $N_c/N_f=x$ and $g^2N_c$ fixed.  
Another distinct large $N_c$ limit with quarks in the two-index anti-symmetric representation was also suggested by 't Hooft \cite{'tHooft:1973jz} and was further explored by Corrigan and Ramond \cite{Corrigan:1979xf}. This limit has generated considerable interest of late \cite{PhysRevD.89.054018, Bolognesi:2009vm, Armoni:2013kpa, DeGrand:2013yja}.
 The two-index anti-symmetric representation labels each quark by two fundamental color labels with $q_{a b}=-q_{b a}$.
The large $N_c$ limit of this theory differs significantly from the standard 't Hooft large $N_c$ limit with quarks in the fundamental, QCD(F), since in QCD(AS) quark loops are not suppressed compared to the gluon loops. The phenomenology of this limit was explored by Kiritsis and Papavassiliou \cite{Kiritsis:1989ge} and baryons in this limit were considered in detail in refs.\cite{Bolognesi:2006ws,Cherman:2006iy,Cohen:2009wm,Cherman:2009fh}.

Note that if one's interest is in the formal structure of gauge theories as opposed to direct application to the phenomenology of QCD, there are other representations for quarks which may be of interest. One obvious one is the adjoint representation, QCD(Adj) in which quarks transform in the same way as do gluons. QCD(Adj) has quarks in what is effectively a two-index representation with one index transforming like a fundamental color and the other as an anti-fundamental. Another representation of interest is the two-index symmetric QCD(S) in which each quark is labeled by two fundamental color labels with $q_{a b}=q_{b a}$.

The two-index theories in the large $N_c$ limit have some very interesting formal properties. Of  particular importance is the emergence of equivalences between the theories at large $N_c$. That is QCD(AS), QCD(S) and QCD(Adj) share a ``common sector'' of operators for which all observables in the sector are identical for the three theories up to corrections which vanish as $N_c \rightarrow \infty$ \cite{PhysRevD.71.045015,PhysRevD.74.105019}. As stressed by \cite{PhysRevD.71.045015}, 
 this is particularly important for the case where there is only  one flavor of quark and it is massless. In this case QCD(Adj) is simply super Yang-Mills. Thus, certain exact results which can be obtained due to the strong symmetry constraints in SYM are also valid at large $N_c$ for the non-supersymmetric theories of QCD(AS) and QCD(S).  

Since large $N_c$ theories with two-index representations are so interesting, it seems natural to consider theories with quarks in representations with three or more indices. To date, such theories have not been systematically studied at large $N_c$. One obvious reason for this is that in 3+1 space-time dimensions, such theories are sick. They lack asymptotic freedom and are thus believed to not be ultraviolet complete. Thus, by themselves, they are not well-defined as theories. However, this does not mean that all theories of this sort lack meaning. One can consider these theories in lower space-time dimensions (either $2+1$ or $1+1$) where the theories are expected to be UV complete and therefore perfectly well-defined. QCD has been explored in lower space-time dimensions in the past \cite{Bagan:2000dw, Leigh:2004ja, Hansson:1997fz, Korchemsky:1996kh, Li:1994kx, Gies:1996ki}.

In this paper we investigate the large $N_c$ behavior of theories with quarks in higher-dimensional representations. Higher dimensional in this context means that the dimension of the representation, $R$, scales with $N_c$ at large $N_c$  as 
\begin{equation}
R \sim N_c^a  \;\;\;{\rm with}  \;\;\; a \ge 3 \; .
\end{equation}
One class of such representations is the one with its Young tableau composed of $a$ boxes, with $a$ independent of $N_c$ and greater than or equal to $3$. These are representations that can be constructed by combining $a$ fundamental colors.  

More generally, we consider representations associated with a Young tableau consisting of $n$ columns each with a length of $a_i$ (where $i$ runs from $1$ to $n$) and $m$ columns each with a length of $N_c-b_j$ (where $j$ runs from $1$ to $m$). Such representations scale at large $N_c$ as $N_c^a$, with
\begin{equation}
a= \sum_{i=1}^n a_i  + \sum_{j=1}^m b_j  \; .
\end{equation}
One can construct such representations by combining $\sum_i a_i$ fundamental indices with $\sum_{j=1}^m b_j$ anti-fundamental ones in such a way that no pair of fundamental and anti-fundamental colors reduces to a singlet. 

Clearly such theories do not describe the underlying dynamics of nature. Indeed, in a mathematical sense, such theories presumably do not exist except in $2+1$ space-time dimensions or fewer. However, it remains of interest to study these theories because they may help give insight into some of the major issues of gauge theory, including perhaps the nature of confinement. Much of the analysis in this paper will be general. However, at times it will be useful to illustrate things using a specific example. In these cases we will focus on the three-index anti-symmetric representation.  

In doing the analysis it is important to be very clear about precisely what is being held fixed as $N_c \rightarrow\infty$. Following standard analysis we study correlation functions in which the external momenta (and quark masses) are held fixed as $N_c \rightarrow \infty$. The scaling of the coupling constant with $N_c$ turns out to be nontrivial. In the next section, we discuss the scaling of the coupling constant with $N_c$. The $\beta$ function will be discussed in the following section.  The key issue there is assuring the existence of an asymptotically free regime. It turns out the theory is conformal in the IR \cite{Poppitz:2009kz} rather than confining if the theory has massless quarks but is confining if the quarks are massive. Following this is a section on correlation functions for local color-singlet sources. Both quark bilinear sources and gluonic sources are considered. The role of confinement is discussed in the next section. A central issue is that at large $N_c$ for the case of massive quarks, the scale of confinement is parametrically suppressed in powers of $1/N_c$ relative to the quark mass and the dynamical scale associated with asymptotic freedom. Finally, we discuss the results and conclude. In the discussion we note that the behavior of these theories is qualitatively similar to theories with fixed $N_c$ and many flavors of quarks in any representation, including the fundamental.

\section{Scaling of the coupling constant}\label{cc}

In the standard large $N_c$ limit of 't Hooft with quarks in the fundamental representation, the number of colors goes to infinity while the coupling constant, $g$ goes to zero with $g^2 N_c$ held fixed \cite{'tHooft:1973jz}. 
\begin{figure*}

\begin{tabular}{ccc}
\subfloat[]
{\label{fig:f0}
\includegraphics[width=0.3\textwidth]{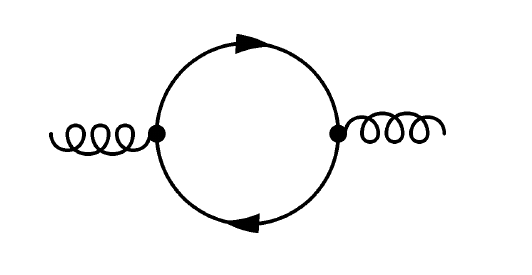}}
&
\subfloat[]
{\label{fig:f1}
\includegraphics[width=0.3\textwidth]{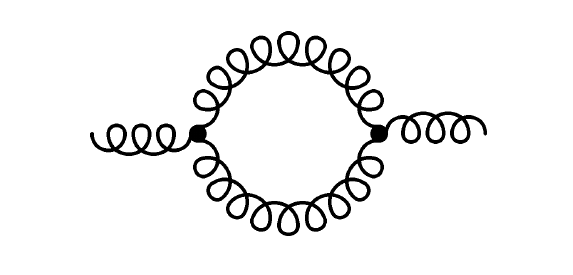}}
\\
\subfloat[]
{\label{fig:f2}
\includegraphics[width=0.3\textwidth]{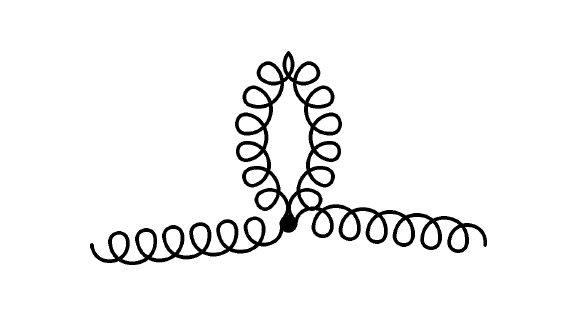}}
&
\subfloat[]
{\label{fig:f3}
\includegraphics[width=0.3\textwidth]{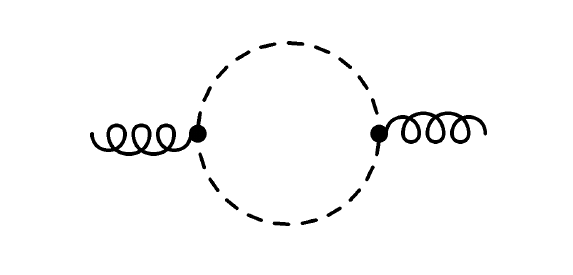}}
\end{tabular}
\caption{\label{fig:gluon_polarization} Diagrams contributing to gluon polarization at one loop. Curly lines stand for gluons, dashed lines for ghost fields and solid lines for quarks.}
\end{figure*}

The simplest way to motivate this is via the study of the gluon propagator. If one wishes the gluon propagator to have a smooth and non-trivial large $N_c$ limit, then the gluon polarization tensor must be held fixed as $N_c \rightarrow \infty$. To proceed further, look at the simplest contribution to the gluon polarization---namely, one-loop diagrams. As seen in Fig.~\ref{fig:gluon_polarization}, there are four possible types of one-loop diagrams: a quark loop (a), two types of gluon loops (b) and (c), and a ghost loop (d). Using standard counting rules, it is easy to see that in the conventional `t Hooft large $N_c$ limit with quarks in the fundamental, the single quark loop contribution to the polarization from diagram (a) is proportional to $g^2$ while the contributions from the gluon and the ghost loops in (b), (c) and (d) are proportional to $g^2 N_c$. Assuming that the contributions of order $g^2 N_c$ do not cancel out, one concludes that the quantity $g^2 N_c$ must remain finite as $N_c$ goes to infinity. Of course, the statement that  $g^2 N_c$ must remain finite, is not the same as it remains fixed---it could, in principle, go to zero as $N_c$ goes to infinity. However, it is straightforward to show \cite{Witten:1979kh} that keeping $g^2 N_c$ fixed leads to a non-trivial and self-consistent theory. Note that the result that quark loops are suppressed compared to gluon loops seen in the gluon propagator, turns out to be general.  

Now, let's consider what happens in a theory in which quarks are in a higher representation, with a dimension scaling as $N_c^a$. Again, let us motivate the scaling rules by looking at one-loop contributions to the gluon proagator as seen in Fig.~\ref{fig:gluon_polarization}. The gluon and ghost loops contribution from diagrams (b), (c) and (d) to the gluon polarization scale as $g^2 N_c$ as before. However, the quark loop of diagram (a) yields a contribution to the polarization which scales as  $g^2 N_c^{a-1}$. For $a > 2$, the quark loop scales more rapidly than the gluon loop. Thus to keep the gluon polarization finite one should take the scaling to be:
\begin{equation}
\label{Eq:scale}
\begin{split}
&N_c \rightarrow \infty\\
&g \rightarrow 0\\
&g^2 N_c^{a-1} \; \; \; {\rm fixed}  \; \; \; \;  {\rm for} \; a>2\\
&g^2 N_c  \; \; \; {\rm fixed}  \; \; \; \;  \; \; \; \; {\rm for} \; a \le 2 \, .
\end{split} 
\end{equation}

Note, that the scalings are very different for $a>2$ and $a<2$. As will be shown below, this reflects qualitatively different physics in the two regimes. In the $a<2$ regime, the dynamics is dominated by gluons and the effects of quarks are suppressed. However for $a>2$ the dynamics is dominated by quarks and gluons play a subsidiary role. For the case of $a=2$, quarks and gluons both contribute at leading order.  

\begin{figure*}
\includegraphics[width=0.9\textwidth]{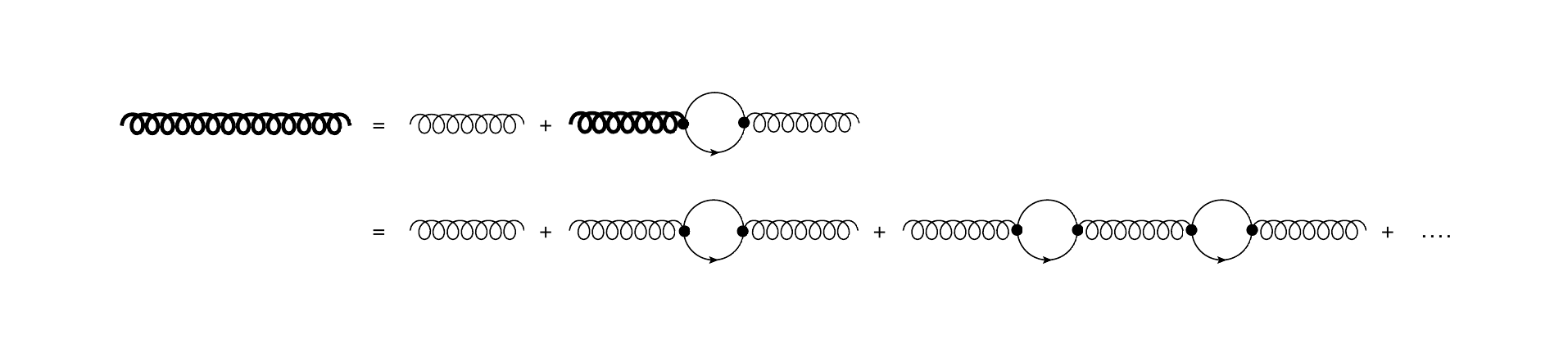}
\caption{\label{Fig:SD} The bold gluon propagator indicates the {\it resummed} one in which the polarization is {\it resummed} to all orders.}
\end{figure*}

It is straightforward to show that this scaling is self-consistent for the case of $a>2$. Firstly, note that since the gluon polarization has been constructed to scale as $N_c^0$ at leading order in the $1/N_c$ expansion, in considering the full class of leading-order diagrams, it is efficient to sum polarization insertions to all orders leading to a renormalized propagator as in Fig.~\ref{Fig:SD}. This is efficient since in considering all diagrams which contribute at leading order, one can now use this {\it resummed} propagator everywhere and exclude explicit quark loop contributions to the polarization everywhere. This allows one to treat infinite classes of leading diagrams at once.

Next, consider a leading-order diagram and ask what happens to the $N_c$ counting if one adds an extra {\it resummed} gluon line in it. As in the standard case of QCD with quarks in the fundamental, the addition of a gluon, will contribute an additional two factors of $g$ and at most a combinatoric factor of $N_c$ (if the gluon is planar). Thus, the addition of an extra internal gluon line to a diagram will contribute a factor scaling as $g^2 N_c$ or less. However for the case of higher representations, Eq.~(\ref{Eq:scale}) implies that the combination $g^2 N_c^{a-1}$ is held fixed as $N_c \rightarrow \infty$ for $a>2$. Rewriting $g^2 N_c$ as $(g^2 N_c^{a-1})  N_c^{2-a}$, one sees that the addition of an extra {\it resummed} gluon line to a diagram characteristically reduces the scaling of the diagram by a factor which scales as $N_c^{2-a}$.  Similarly, removing a {\it resummed} gluon line from a diagram increases the scaling of the diagram by a factor which scales at least as $N_c^{2-a}$.

The upshot of this scaling is that for quarks in higher representations ($a>2$), the maximum $N_c$ scaling of any class of diagrams necessarily consists of diagrams with the smallest number of {\it resummed} gluon lines consistent with the type of diagram under study. Thus, for example, the leading vacuum amplitude (i.e. a zero point function) of quark bilinear sources will be from a single quark loop and will scale as $N_c^a$. Similarly, the leading contribution to correlation functions of any number of quark bilinear sources, will consist of a single quark loop embellished by the sources and will scale as $N_c^a$. On the other hand, the leading-order correlation functions for sources which couple to glue such as ${\rm  Tr} \left ( F_{\mu \nu}F^{\mu \nu} \right )$ will consist of a single loop of the {\it resummed} gluon embellished by the sources and will scale as $N_c^2$.  

At first sight the scaling rules may appear to be trivial. The leading-order contribution to correlators of the quark bilinear sources is simply a single quark loop, as one would have in a weak coupling. However, the theory differs markedly from the case of a pure weak coupling theory. Note, that the leading corrections involve the {\it resummed} gluon propagator and as such contain the coupling constant to all orders. Similarly, the leading order correlation functions for color-singlet sources coupling to gluons again involve the {\it resummed} gluon propagator and as such contain the coupling constant to all orders.  

\section{Renormalization group flow}  \label{CC2}

As noted in the Introduction, large $N_c$ gauge theories with quarks in higher representation are not asymptotically free in $3+1$ space-time dimensions. Accordingly, we consider theories in fewer space-time dimensions, either $2+1$ or $1+1$. The gauge coupling $g$, while dimensionless in theories in $3+1$ space-time dimensions, is dimensionful in lower dimensions: in $d$ space-time dimensions the coupling constant has a dimension of $\frac{4-d}{2}$. It is useful to define a dimensionless coupling $\tilde{g}$, which we do by introducing an arbitrary renormalization scale $\mu$:
\begin{equation}
g= \tilde{g}\mu^{\frac{4-d}{2}} \; .
\end{equation}

The $\beta$ function is defined as $\beta (\tilde{g})=\frac{\partial \tilde{g}}{\partial \log (\mu )}$. It has two contributions: one coming from the explicit dependence of $\mu$ in the naive scaling dimension of the coupling and the other coming from quantum loops. For simplicity we first consider the case in which all quarks are massless. Thus, the $\beta$ function is given by 
\begin{equation}
\begin{split}
\beta(\tilde{g}) & =\tilde{g} \frac{d-4}{2}+\tilde{\beta}(\tilde{g}) \\ \text{where}\\
\tilde{\beta}(\tilde{g}) &= \tilde{g} \left ( b_1 \tilde{g}^2 +  b_2 \tilde{g}^4 + b_3 \tilde{g}^3   +\ldots       \right )\\
&= \tilde{g} \left (\frac{ b_1}{N_c^{a-1}  }(\tilde{g}^2 N_c^{a-1}) +   \frac{ b_2}{(N_c^{a-1} )^2 }(\tilde{g}^2 N_c^{a-1})^2   + \frac{ b_3}{(N_c^{a-1} )^3 }(\tilde{g}^2 N_c^{a-1})^3  + \ldots \right ).
\end{split}
\end{equation}
The form for $\tilde{\beta}$ follows from a loop expansion with the coefficient $b_i$ associated with $i$ loops. The second form for $\tilde{\beta}(\tilde{g})$ is introduced to emphasize the scaling behavior of Eq.~(\ref{Eq:scale}) for the case of quarks in higher representations.

Note that the loop contributions in the $\beta$ function can involve either quarks or gluons and ghosts. For higher representations, quark loops yield a contribution proportional $N_c^{a-1}$ while gluon or ghost loops yield contributions proportional to $N_c$. The factors of $b_i/ (N_c^{a-1} )^i$ thus, will go to zero as $N_c \rightarrow\infty$ except for contributions in which {\it all} of the loops are quark loops. However, the structure of the quantum loops which yield renormalization group equation implies that the only contribution in which all loops are quark loops is at one loop. Thus at large $N_c$, the renormalization group equation assumes the form
\begin{equation}
\beta(\tilde{g})  =\tilde{g} \left( \frac{d-4}{2}+ \frac{c_d}{(a-1)!} \, N_F (\tilde{g}^2 N_c^{a-1}) \right )
\label{Eq:beta}\end{equation}
where $c_d$ is a numerical constant which depends on the dimensions of space-time $c_d=\frac{1}{32}$ in $2+1$ dimensions and is $\frac{1}{2\pi}$ in $1+1$ dimensions. $N_F$ is the number of (massless) flavors. It is straightforward to solve the differential equation for the large $N_c$ coupling as a function of $\mu$:
\begin{equation}
g^2(\mu) \frac{N_c^{a-1}}{(a-1)!} = \Lambda^{4-d} \left (\frac{4-d}{2 c_d N_F} \right ) \frac{1}{1 + \left(\frac{\Lambda}{\mu} \right )^{4-d} }
\label{Eq:renormsol}\end{equation}
where $\Lambda$, the natural scale of the theory, is fixed from the initial condition of the differential equation. As expected, $g^2 N_c^{a-1}$ or $g^2 \frac{N_c^{a-1}}{(a-1)!}$ stays fixed at large $N_c$ and the theory encodes asymptotic freedom---while $g(\mu)$ asymptotes to a fixed value at large $\mu$, the dimensionless coupling $\tilde{g}(\mu)$ goes to zero.  

In the infrared, $g^2(\mu) \frac{N_c^{a-1}}{(a-1)!} $ asymptotes to $\frac{4-d}{2 c_d \, N_F} \mu^{4-d}$. For dimensionless coupling $\tilde{g}$, this corresponds to asymptotic behavior in the infrared corresponding to ${\tilde g(\mu)}^2 \frac{N_c^{a-1}}{(a-1)!}$ going to $\frac{4-d}{2 c_d N_F}$. This is easy to understand: from Eq.~(\ref{Eq:beta}), it can be seen that the beta function vanishes for $\tilde{g}^2 \frac{N_c^{a-1}}{(a-1)!}= \frac{4-d}{2 c_d N_F} $. Thus, the theory approaches a fixed point in the infrared: it becomes conformally invariant. Unlike pure YM, the theory has no mass gap. 
Note, of course that as written Eq.~(\ref{Eq:renormsol}) is valid only for the case of massless quarks, since in the massless case the one-quark loop contribution to the $\beta$ function depends on $\mu$ only implicitly through $g$. Moreover, for massless quarks,  the result is exact at large $N_c$. However, when non-zero quark masses are taken into account, the running of the coupling depends on the ratio of $\mu$ to $m_q$. Note that Eq.~(\ref{Eq:renormsol}) remains valid for $\mu \gg m_q$ since in that case the quark mass is irrelevant and behavior is that of the massless case. For the purposes of verifying asymptotic freedom this is sufficient. On the other hand for $\mu \ll m_q$ the quarks are frozen out. Since the quark loop is the only source of running at large $N_c$, one expects that $\tilde{g}$, the dimensionless coupling, follows Eq.~(\ref{Eq:renormsol}) at large $\mu$ but it slows down as $\mu$ approaches the regime of $m_q$ and stops asymptotically as $\mu$ gets much smaller than $m_q$. If there are multiple flavors of quarks with different masses, then one expects the form of Eq.~(\ref{Eq:renormsol}) to hold for $\mu$ well away from any of the quark masses with $N_f$ equal to the number of active quarks (quarks with masses well below $\mu$). The values of $\Lambda$ used in Eq.~(\ref{Eq:renormsol}) will differ in the various regions; they will be fixed by the property that the coupling constant as a function of $\mu$ needs to smoothly connect from below the threshold in which a quark is inactive to the one above it.

\section{Correlation functions}\label{sec:corr}

Correlation functions in these theories are of interest. By ``glueball-glueball'' correlation function, we mean the correlator for the purely gluonic local color singlet source $ \sum_a F^a_{\mu \nu}  F^{a \; \mu \nu}$ where $a$ represents color. Similarly the ``meson-meson'' correlation functions are the correlators for quark bilinear sources. As it happens, the glueball-glueball  and the meson-meson correlation functions are both exactly calculable in the large $N_c$ limit. This is because as was shown in the previous section, the addition of an extra {\it resummed} gluon line to a diagram characteristically reduces the scaling of the diagram by a factor which scales as $N_c^{2-a}$. Thus, the leading diagrams are those with the fewest number of {\it resummed} gluon lines.

In  the case of  the meson-meson correlation function, the one-loop diagram is of the order $N_c^{a}$, and any higher loop diagram is suppressed by powers of $N_c^{2-a}$. 
For concreteness we illustrate the general issues associated with meson correlators in the case of scalar sources for one flavor of massless quarks in the three-index anti-symmetric representation for QCD in $2+1$ space-time dimensions. A quark in this theory is of the form $q_{i j k}$ where $i,j,k$ run from $1$ to $N_c$ and where $q_{i j k}= - q_{j i k}=- q_{k j i}$. The scalar meson correlation function in general is given by
\begin{equation}
\Pi_{\rm meson}(k^2)=\int \langle {\rm T}  \left [ \bar{q}^{ijk}q_{ijk}\left(x\right)\bar{q}^{lmn}q_{lmn}\left(0\right) \right] \rangle e^{ikx}d^3x \, ,
\end{equation}
where T indicates time-ordered product and the color indices $i, j, k, l, m, n$ are summed. At leading order in $N_c$ we have
\bea
\Pi_{\rm meson}^{\rm LO}(k^2) &=& \frac{N_c^3}{3!}\int \frac{d^3q}{\left(2\pi\right)^3} \text{Trace}\left[\left(\frac{i\left(\gamma_{\mu} q^{\mu}+m_q\right)}{q^2-m_q^2}\right)\left(\frac{i\left(\gamma_{\mu}k^{\mu}+\gamma_{\mu} q^{\mu}+m_q\right)}{(q+k)^2-m_q^2}\right)\right]\nn\\
& =& \frac{-iN_c^{3}}{6\pi}\int_0^1\sqrt{m_q^2-k^2x(1-x)}dx  + {\rm const} = \frac{-iN_c^{3}}{6\pi} \left (  \frac{m_q}{2} -  \frac{ \left (k^2 -4 m_q^2  \right )  \coth^{-1} \left( \frac{2 m_q}{\sqrt{k^2}} \right ) }{4 \sqrt{k^2}} \right )  + {\rm const} \; ,
\label{Eq:mm}\eea 
where the constant arises due to the need to renormalize the (divergent) composite operator.

The general structure of Eq.~\eqref{Eq:mm} holds for scalar meson correlators for theories in $2+1$ space-time dimensions with quarks in any higher representation and with any number of degenerate flavors. The only modification is a different overall factor. A few simple comments about this structure are in order. The first is that after neglecting the additive constant, $-i \Pi_{\rm meson}^{\rm LO}(k^2)$ is purely real for $p^2 < 4m_q^2$. It develops an imaginary part at $p^2 = 4m_q^2$, which corresponds to the threshold for unconfined $\overline{q}$-$q$ pair production. This may be a bit of a surprise: while the massless theory is conformal in the IR, the theory with massive quarks is confining. However, as will be discussed in sec.~\ref{Conf}, the scale of confinement is parametrically suppressed in $N_c$ and thus, the correlator is accurately described by the expression for unconfined quarks except right in the immediate vicinity of the would-be threshold. The massless limit of the correlator is of interest: $\Pi_{\rm meson}^{\rm LO}(k^2) \rightarrow \frac{-i 4 N_c^{3}}{3 \pi} \sqrt{- k^2} + {\rm const}$.



%

The glueball correlation functions at leading order are also straightforward. One simply calculates the one-loop correlation function using the dressed ({\it i.e.} {\it resummed}) gluon propagator from Fig.~\ref{Fig:SD}. The leading order diagrams go as $N_c^2$. Corrections associated with diagrams with additional gluon lines are suppressed by powers of $N_c^{2-a}$.  
The first step is to compute the dressed gluon propagator. It is worth noting that the dressed propagator involves renormalization and must be specified at a scale. In order to keep the calculation consistent with that of the $\beta$ function, it is natural to set the scale for the dressed propagator to be the same as the scale of the couplings used in the bubble sum. The result is particularly simple in the massless case:
\begin{equation} 
\begin{split}
D^R_{\mu \nu}(q^2,\mu) & =\frac{g_{\mu\nu}}{(q^2+i\epsilon) \left (1+2c_d N_f \tilde{g}(\mu)^2\frac{N_c^{a-1}}{(a-1)!} (i\frac{\mu}{\sqrt{q^2}}-1)\right )}\\
 &=  \frac{g_{\mu\nu}}{(q^2+i\epsilon) \left (1+\frac{i\frac{\mu}{\sqrt{q^2}}-1}{\frac{\mu}{\lambda}+1}\right)} .  \end{split}
\end{equation}
The expression for the {\it resummed} propagator for the case of massive quarks is significantly  more complicated but it too can be expressed in closed form.

This dressed propagator leads directly to the scalar glueball-glueball correlation function,
\bea
\label{glue}
\Pi_{\rm glue} \left(p\right)=\int \frac{d^3k}{\left(2\pi\right)^3} D^{\rm R}_{\alpha \beta} \left(k^2, \mu \right) D^{{\rm R} \, \alpha \beta} \left((p+k)^2 ,\mu \right).
\eea
There does not appear to be any closed-form expression for this integral, even in the massless case.  However, the integral can be evaluated numerically in a straightforward way.

\section{Confinement}\label{Conf}

The theory with massive quarks---unlike the case of massless ones---is confining. The reason is simple: the infrared physics is dominated by gluons at scales well below the quark mass since the quarks are frozen out and the gluodynamics is confining, not conformal. However, this gives rise to an apparent puzzle: if the theory is confining, then Eq.~(\ref{Eq:mm}) may seem problematic. After all, this expression is nothing but the correlation function for noninteracting quarks. How can a confining theory yield the correlator for unconfined quarks ?

Actually, there is a very natural way for this to occur. To understand this, it is useful to recall what happens to correlators of quark bilinears with four-momenta that are large compared to $\Lambda_{\rm QCD}$ in ordinary QCD in $3+1$ dimensions. As is well known, such correlators are accurately described by free-field correlation functions and become increasingly well described this way as the four-momentum increases. Now, this is usually understood as resulting from asymptotic freedom---the theory becomes increasingly weakly coupled and for the purposes of describing the correlator the quarks act, to good approximation, as though they are free and unconfined. This understanding is correct so far as it goes.  

However, there is an alternative way to think about the behavior of the large $q^2$ correlator in QCD which sheds light upon the present problem. From general principles \cite{Peskin}, the correlator can be written in Kallen-Lehman representation:
\begin{equation}
\pi(q^2) = \int d s \frac{\rho(s)}{q^2 -s + i\eps}
\end{equation}
where $\rho(s)$ is the square of the amplitude for the quark bilinear source to create a  physical state with mass of $\sqrt{s}$. In order, for the large $q^2$ correlators to be accurately described by the free quark-antiquark result, $\rho(s)$ must also be accurately described by the free theory result at large $s$. But $\rho(s)$ describes the amplitude for creating {\it physical} states and the physical states are made of hadrons, with quarks confined in them. Somehow the spectral density, although actually composed of contributions from physical multi-hadron states, manages to simulate the behavior of a free quark-antiquark pair for sufficiently large $s$. A necessary condition on the regime where this happens is that $s$ should be much larger than the confinement scale for the theory--i.e. the scale that controls the detailed dynamics of the confined hadronic state. The reason for this is simply that in the regime of interest, the spectral function is a smooth function reflecting the phase  space for the would-be free quark-antiquark pair. Thus, the spectral function cannot  be sensitive to the details of the individual confined hadrons which actually compose the state.  This will happen only if there is enough phase space available that the system averages over all of the detailed physics of the individual hadrons at the confinement scale.  

Of course, in ordinary QCD in $3+1$ dimensions, these two perspectives on the correlator at large $q^2$ are complementary. They are different ways to think about the problem and deal with different aspects. However, the two perspectives are completely consistent with each other: in QCD, there is essentially only one scale---$\Lambda_{\rm QCD}$---and it sets both the scale at which asymptotic freedom sets in and the scale where confinement begins. Thus, when $q^2 \gg \Lambda_{\rm QCD}$ one expects that asymptotic freedom forces the correlator to look like the free field one {\it and} in the same regime one expects the spectral function to be insensitive to the confinement dynamics enabling the spectral function to do so.

The question of interest here is the behavior of large $N_c$ QCD with quarks in higher representations and lower spatial dimensions. For these models, the behavior is a bit more subtle.  The key thing, which  we will show below, is that unlike for the case of ordinary QCD, the scale which controls the asymptotic behavior of the coupling is parametrically well separated from the confinement scale. In particular, the ratio of the confinement scale, $\Lambda_{\rm conf}$ to  $\Lambda$ scales as
\begin{equation}
\frac{\Lambda_{\rm conf}}{\Lambda} \sim N_c^{\frac{2-a}{4-d}} \; . 
\label{Eq:confscale}\end{equation}
Thus, for example  a three index representation in $2+1$ space-time dimensions, $\Lambda_{\rm conf}/\Lambda$ scales as $1/N_c$. Since the analysis done in sec. \ref{CC2} was in the limit of $N_c$ going to infinity with masses, external momenta and $\Lambda$ held fixed, the  regime studied implicitly had $q \gg \Lambda_{\rm conf}$. Given this scaling, it is perfectly understandable why the dynamics of confinement do not play a role in the meson-meson correlator: one is simply working at a scale well above the confinement scale even though it is not well above $\Lambda$.  

It is easy to derive the scaling in Eq.~(\ref{Eq:confscale}). Let us return to the analysis of sec. \ref{cc} and for simplicity assume non-zero quark masses with either a single flavor of quarks or degenerate flavors so that there is only one quark mass in the problem. It was argued in sec. \ref{cc} that Eq.~(\ref{Eq:renormsol}) holds for $q \gg m_q$. It was stated that the running of the dimensionless coupling slows down as $q$ becomes comparable to $m_q$ and stops asymptotically as $q$ gets much smaller than $m_q$. This is correct as far as it goes. However, this analysis holds only for $q$ of order $N_c^0$ when the leading order dynamics dominates. For sufficiently small $q$ one cannot neglect the subleading effect in $1/N_c$ associated with gluon exchange and running begins again. We will see that ``sufficiently small" means a $q$ which is parametrically of the order $N_c^{\frac{2-a}{4-d}}$ (and thus is pushed to zero in the limit of large $N_c$).

Let us consider the value of the coupling at renormalization scale $\mu$, $g(\mu)$ in a regime in which $\mu$ is both much smaller than $m_q$, but also of order $N_c^0$.  If  such a regime is approached from above, Eq.~(\ref{Eq:renormsol}) is accurate for $\mu \gg m_q$. The running slows down and stops as $\mu$ approaches and then drops well below $m_q$. A very crude estimate of the value of $g$ in the regime under consideration would be to assume that Eq.~(\ref{Eq:renormsol}) hold for $\mu > m_q$ and then running stops immediately when $\mu$ hits $m_q$. The actual value will differ from this crude estimate due to the running in the regime $\mu \sim m_q$. It is clear that such running can lead to a correction to the crude estimate by a factor of order $N_c^0$ since the running shuts off over a region of order $N_c^0$. Thus the coupling in the regime of interest is given by 
\begin{equation}
g^2 = f(a-1)!N_c^{1-a}\Lambda^{4-d} \left (\frac{4-d}{2 c \,  N_F} \right ) \frac{1}{1+ \left(\frac{\Lambda}{m_q} \right )^{4-d} }
\label{Eq:gabove}
\end{equation}
where $f$ is the correction factor which accounts for running with $\mu\sim m_q$; $f$ can be explicitly computed but its precise value is not of concern here.

Now let us consider what happens if we approach from below the regime in which $\mu$ is both much smaller than $m_q$, but also of order $N_c^0$. Let us begin running with arbitrarily small $\mu$ so that $\mu$ is not of the order $N_c^0$ and ask what happens as it grows towards $N_c^0$. In this case, the role of the quark in the renormalization group flow can be neglected, but the role of the gluons cannot. Thus, the theory runs the same way as a pure Yang-Mills theory does.  In a pure Yang-Mills theory the leading order contributions in $1/N_c$ are planar and always have the coupling in the combination $g^2 N_c$. Thus, the form of the scaling at leading order  is  
\begin{equation}
g^2  (\mu) = N_c^{-1} \Lambda_{\rm conf}^{4-d} \, \, h\left(\frac{\mu}{\Lambda_{\rm conf}} \right)
\end{equation} 
where $\Lambda_{\rm conf}$ is the confinement scale associated with the Yang-Mills theory and $h$ is a function characterizing the scaling. Note that the lower dimensional Yang-Mills theory is asymptotically free in the sense that $g^2/\mu^{4-d}$ goes down with increasing $\mu$. It is easy to show from the renormalization group equation that the dimensionless coupling asymptotes to a constant. Thus the function $h$ has the property that the limit  of $h(x)$ as $x$ goes to infinity is a finite, nonzero value which we denote $h_\infty$.  Thus as one approaches the regime of interest from below, the running stops and one obtains
\begin{equation}
g^2  = N_c^{-1} \Lambda_{\rm conf}^{4-d} \, \, h_{\rm \infty} \\ .
\label{Eq:gbelow} \end{equation} 
Equating Eqs.~(\ref{Eq:gabove}) and (\ref{Eq:gbelow}) yields Eq.~(\ref{Eq:confscale}).

\section{Discussion}\label{sec:conclusion}

The large $N_c$ gauge theories discussed in this paper are very different from the typical large $N_c$ gauge theories. In the regime where external momenta and quark masses are taken to be of order $N_c^0$, the dynamics is dominated by the quark loops and the confining dynamics associated with gluodynamics is irrelevant. This  means that the large $N_c$, $\beta$ function is exactly calculable; it is given by  Eq.~(\ref{Eq:beta}) for the case of massless quarks. In the massless quark case, the theory becomes conformal in the infrared. In the case of nonzero quark masses, the theory is confining. However, the confining scale is parametrically well separated from the scale $\Lambda$ which parameterizes the scaling of the coupling in the ultraviolet by an amount given by Eq.~(\ref{Eq:confscale}). Correlation functions for color-singlet quark bilinear and gluon bilinear sources are easily computed in this limit.

It is worth noting that the behavior seen in this version of the large $N_c$ limit is qualitatively similar to gauge theories in other regimes in which the quark loops dominate. Thus, for example, they will behave quite similar to gauge theories with fixed $N_c$ and many degenerate flavors of quark in any representation including the fundamental. As in the case of large $N_c$ with quarks in the higher representations, such theories are not asymptotically free in $3+1$ D as the quark loops dominate the beta function. In order to have a smooth $N_f$ limit that keeps the gluon polarization finite we need to hold $g^2N_f$ fixed as $N_f\rightarrow\infty$. Once again if we go to lower space-time dimensions, in the absence of quark mass the beta function looks like
\begin{equation}
\beta(\tilde{g})  =\tilde{g} \left( \frac{d-4}{2}+ \tilde{c} \,  (\tilde{g}^2 N_f) \right )
\label{Eq:beta2}\end{equation}
where the constant $\tilde{c}$ depends on dimensions of space-time and number of colors $N_c$ which is finite in this case. The form of this beta function is identical to \eqref{Eq:beta} and as before we approach a non-interacting theory as we go to higher and higher energies and achieve conformality in the infrared. Introduction of massive quarks gives rise to confinement as before with the confinement scale parametrically separated from the ultraviolet scale. The factor separating the two scales is given by $N_f^{\frac{1}{4-d}}$.
\section*{Acknowledgements}
The authors would like to thank Aleksey Cherman for insightful discussions. This work was supported by the U.S. Department of Energy through grant number DEFG02-93ER-40762.

\bibliographystyle{unsrt}
\bibliography{threeindex}
\end{document}